\documentclass[amsmath,amssymb,
 aps,pre,onecolumn,superscriptaddress,floatfix
]{revtex4-2}
\usepackage{graphicx}
\usepackage{dcolumn}
\usepackage{bm}
\usepackage{amsmath}
\usepackage{mathtools}
\usepackage{xcolor}
\usepackage[export]{adjustbox}
\newcommand{\resub}[1]{\textcolor{black}{#1}}
\begin{document}
\title{Microscopic modifications of equilibrium probabilities due to \resub{non-conservative} perturbations \resub{with applications to anharmonic systems}}
\author{
Dino Osmanovi{\'{c}}
}
\affiliation{Department of Mechanical and Aeronautical Engineering, University of California, Los Angeles, Los Angeles, CA, USA}%
\email{osmanovic.dino@gmail.com}
\date{\today}

\begin{abstract}
    The standard relationships of statistical mechanics are upended my the presence of active forces. In particular, it is no longer usually possible to simply write down what the stationary probability of a state of such a system will be, as can be done in ordinary statistical mechanics. While exact expressions are possible for harmonic systems, anharmonic systems tend to be much more difficult to treat. In this manuscript, we investigate how the microscopic probability for anharmonic systems is modified in the presence of non-conservative forces. We recount how non-conservative corrections to microscopic probabilities in generic systems can be represented as integrals over Green's function kernels, and that these Green's functions take the form of path integrals. We show how using analytically tractable form of these functions allows us to calculate corrections to the microscopic probability for a generic anharmonic system. These results are compared to active Brownian particles inside a harmonic or anharmonic well. From this it can be shown that accumulation of probability away from its equilibrium minima is a natural effect arising in anharmonic but not harmonic systems. Finally, we extend the microscopic results to study small active polymers with anharmonic backbone potentials. The interplay of the active driving with the anharmonic backbone potential can be used to understand how the end-to-end distance and mode space distributions of anharmonic active polymers scales with active driving. In particular, the polymer modes with the smallest eigenvalues will be affected by non-conservative forces the most.
\end{abstract}

\maketitle
\section{Introduction}

Active systems constitute a class of physical models whose continual consumption of energy requires the use of new theoretical methods and ideas to describe their behavior\cite{ramaswamy_2010}. The presence of energy-consuming processes leads to violations of the conditions necessary for the establishment of equilibrium states. Such systems are of interest as many phenomena belong to this domain; most importantly, this is a defining characteristic of biological systems\cite{Needleman2017}.

Analysis of problems in these areas would be deemed to be related to non-equilibrium physics. However, the analysis of non-equilibrium systems is complicated by the fact that "non-equilibrium" denotes a variety of different systems, each of which has its own peculiarities. For example, glasses are not in equilibrium\cite{Stillinger2013}, though intuitively appear as a different ``kind" of non-equilibrium to a living cell. Other examples of non-equilibrium could include maintaining two different parts of a system at two different temperatures, leading to continuous heat transfer from the hotter part of the system to the cooler one\cite{Netz2020}. Lastly, ``classical" active matter (as described above) is often thought of as using consumption of energy for directed motion\cite{Cates2015}, though its definition can also include consumption of energy to drive non-directed processes\cite{Weber_2019}.

Focusing on problems which have relevance for biology, the lack of any unifying description to such systems has necessitated a variety of methods for describing the non-equilibrium stationary states in active systems. A variety of active problems are well described through field theoretic methods, where analysis of the symmetries of the problem allow construction of functionals that can give mean-field behavior \cite{Cates2019}. Alternatively, one can describe all the microscopic degrees of freedom, however, such a level of description usually requires coarse grained simulations, given the lack of useful expressions for the microscopic distributions over states at non-equilibrium, nevertheless, a variety of different systems can be studied in such a manner, including flocking behaviors, active Brownian motion and active chemistry\cite{Toner2005,Callegari2019,Osmanovic2019}. More abstractly, stochastic thermodynamics generalizes standard thermodynamic notions towards the analysis of non-equilibrium steady states\cite{Seifert2012}, which can be extended towards the study of active matter \cite{Speck2016}.

Common to all the above is that the standard Boltzmann distribution over states is no longer sufficient to describe these scenarios\cite{Netz2018}. Notably, however, occasionally such scenarios can be described rather simply through renormalization of temperature (so called active (or effective) temperature)\cite{Loi2008,Loi2011,Bechinger2016,DeKarmakar2020,Caprini2021}. One is left with the question of whether any generalized microscopic description can tell us something about the properties of non-equilibrium steady states.

One way to explore this issue is by treating excess forces in these systems as a perturbation\cite{Moyses2015,Amarouchene2019,Mangeat2019,Noh2015,Wedemann2016,Bertin2017}. The perturbations can be represented as integrals over Green's function kernels\cite{Osmanovic2021}, however, the precise form of these kernels is difficult to give in analytic form, other than as the solution to a complicated partial differential equation. This Green's function governs the response of the microstate probability distribution to additional (not necessarily arising from a potential) forces in the system, leading to a new probability of each microstate. As deviations from Boltzmann distributions can be characterized in terms of integrals over this Green's function kernel, which serves then as a microscopic response function, what can we say about this function that governs microscopic responses to non-equilibrium perturbations? Response functions are a canonical way of treating non-equilibrium states\cite{doi:10.1142/p392,DalCengio2019,}, however they are usually thought of in terms of macroscopic rather than microscopic quantities. We note here the caveat that we use the term ``response" rather generally and in a way that is not necessarily related to its classical usage in fluctuation-dissipation theories. We are interested in the function that tells us how the microscopic distribution of states is modified through the presence of non-equilibrium perturbations, as opposed to the dynamical features of the system in the presence of an additional time dependent force.

Prior works have demonstrated interpretable analytic expressions for harmonic systems\cite{Kwon2011}  (systems where the forces are linear in the degrees of freedom) subject to dissipative driving. Anharmonic systems have attracted some analytical attention\cite{Das2018,Kedia2019,Su2022}, though less than harmonic systems, possibly due to the general difficulty involved in treating anharmonic forces.

In this manuscript, we set out to explore how this microscopic distribution over states for non-conservative perturbation can be calculated, for a generic system including anharmonic conservative forces. To this end, we first demonstrate that the Green's function can be written as a path integral over all paths in the equilibrium system and give an interpretation of the expression that arises. The similarity of this path integral to those that arise in quantum theory allows us to derive analytical results for anharmonic systems. We derive these corrections for an active Brownian particle inside an anharmonic trap and compare these results against simulations of the same. We demonstrate that approximate forms of the Green's function kernel can generate correct results even up to quite large values of the activity. We then apply the microscopic expressions that arise from the study of the anharmonic trap to an active polymer with an anharmonic backbone potential.

\section{Results}
\subsection{\resub{Microscopic stationary Green's functions}}

We imagine an equilibrium system by positing a Hamiltonian $H(\mathbf x)$ describing $N$ particles which are also coupled to an external thermal bath at an inverse temperature $\beta=1/k_b T$. At this juncture we introduce a point of nomenclature to help ease confusion. While \textit{Equilibrium} can be said to correspond to a set of states of a system with Hamiltonian $H$, such a system could be instantiated in a particular region of phase space that is not in this ``equilibrium" subset. However, it is (given sufficient time) still possible for such a system to reach equilibrium states. This differs in kind to a system where not all the forces arise from the Hamiltonian, which we refer to as ``non-equilibrium".

  For simplicity, we consider the case where each of these particles are overdamped (i.e. a colloidal system), such that a state contains only the position degrees of freedom of all the particles. In standard statistical mechanics, the probability of a state $\mathbf x$ is proportional to the Boltzmann factor $P(\mathbf x)\sim \exp(-\beta H(\mathbf x))$. However, in the presence of additional forces which cannot be described through potentials $\mathbf F_{nc}(\mathbf x)$, this probability doesn't describe the true probability of a state probability of the system. In a previous manuscript\cite{Osmanovic2021}, we demonstrated through the Smoluchowski equation that the effect of a non-conservative force (to first order) on the microscopic probability appears as:
\begin{equation} \label{eq:nonCforce}
    \log(P(\mathbf x)) \approx - \beta H(\mathbf x) + \int \mathrm d \mathbf x' G(\mathbf x,\mathbf x') \rho(\mathbf F(\mathbf x'))
\end{equation}
where we have suppressed additional constants that would appear on the right hand side. One could also assert ~\eqref{eq:nonCforce} quite generally from linear response theory. The Green's function $G(\mathbf x',\mathbf x)$ quantifies how a perturbation at a point $\mathbf x'$ affects some quantity at point $\mathbf x$. The ``source function'' $\rho(\mathbf F(\mathbf x) )$ is a function of the non-conservative perturbation on the system, but we shall see that it's generally related to the total power dissipated by the process that is being introduced. The precise forms of these functions can be calculated for whatever form of non-conservative perturbation is present. \resub{For example, an equilibrium system subject to some divergence free non-conservative force $\mathbf F_{\text{nc}}(\mathbf x)$ has the following source function:
\begin{equation} \label{eq:source}
\rho(\mathbf x) = \beta^2 \nabla H(\mathbf x).\mathbf F_{\text{nc}}(\mathbf x)
\end{equation}
}
Higher order perturbations (see below) are also integrated over the kernel $G(\mathbf x,\mathbf  x')$, reinforcing its importance in governing how the probability of a state depends on non-conservative forces. The Green's function is also sometimes referred to as a response function. We again note the caveat stated in the introduction over the term ``response". This function has no time dependence, but tell us how the probability at a point $\mathbf x$ ``responds" to a delta function impulse at another point $\mathbf x'$. We can make this clearer by referring to the function as the ``stationary response function". 

It can be demonstrated via analysis of the Smoluchowski equation\cite{Osmanovic2021} that the Green's function for perturbation by a non-conservative force is given by the solution to the following equation:
\begin{equation} \label{eq:pdeGreen}
\left(\nabla^2-\beta \nabla H(\mathbf x) .\nabla\right)G(\mathbf x, \mathbf x') = \delta(\mathbf x- \mathbf x')
\end{equation}
\resub{which applies regardless of the type of perturbation involves coupling to different temperature baths or action by deterministic forces. (See appendix \ref{app:appA} for a derivation of this equation for a multi-bath system)}

As was mentioned previously, the Green's function characterizes how the microscopic probability is modified through the presence of non-potential forces. However, equation \ref{eq:pdeGreen} lacks a very clear intuitive physical interpretation of the important elements of this function. Remarkably, however, equation \ref{eq:pdeGreen} can be solved by a path integral, a derivation which arose in a rather different context (electrostatics)\cite{Gersten1987} but appears nevertheless to apply to non-equilibrium systems. 

The function $G(\mathbf x, \mathbf x')$ can be written as an integral over all paths $\mathbf r(s)$:
\begin{align} \label{eq:Green}
G(\mathbf x, \mathbf x') &= \int_0^{\infty} \!\!\!\!\!\!\mathrm d s \int_{\mathbf r(0)=\mathbf x'}^{\mathbf r(s)=\mathbf x} \!\!\!\!\!\!\mathcal{D}\left[\mathbf r(s)\right]e^{-\int_0^s \mathrm d s' \frac{1}{4}\left(\frac{d \mathbf r(s')}{d s'}\right)^2-\frac{1}{2}\beta\nabla H(\mathbf r(s')).\frac{d \mathbf r(s')}{d s'}+\frac{1}{4}\beta^2\nabla H(\mathbf r(s')).\nabla H(\mathbf r(s'))-\frac{1}{2}\beta\nabla^2 H(\mathbf r(s'))
} \\ \label{eq:GreenL} &= \int_0^{\infty} \!\!\!\!\!\!\mathrm d s \int_{\mathbf r(0)=\mathbf x'}^{\mathbf r(s)=\mathbf x} \!\!\!\!\!\!\mathcal{D}\left[\mathbf r(s)\right]e^{-\int_0^s \mathrm d s' L\left(\frac{d \mathbf r(s')}{d s'},\mathbf r(s'),s' \right)}\end{align}
Which can be rewritten as:
\begin{align}
G(\mathbf x, \mathbf x') &=e^{\frac{\beta}{2}\left(H(x)-H(x')\right) }K(\mathbf x,\mathbf x') \\
K(\mathbf x, \mathbf x') &=\int_0^{\infty} \!\!\!\!\!\!\mathrm d s \int_{\mathbf r(0)=\mathbf x'}^{\mathbf r(s)=\mathbf x} \!\!\!\!\!\!\mathcal{D}\left[\mathbf r(s)\right]e^{-\int_0^s \mathrm d s' \frac{1}{4}\left(\frac{d \mathbf r(s')}{d s'}\right)^2+\frac{1}{4}\beta^2\nabla H(\mathbf r(s')).\nabla H(\mathbf r(s'))-\frac{1}{2}\beta\nabla^2 H(\mathbf r(s'))}
\end{align}
Upon the susbtitution $V(\mathbf r)=\frac{1}{4}\beta^2\nabla H(\mathbf r(s')).\nabla H(\mathbf r(s'))-\frac{1}{2}\beta\nabla^2 H(\mathbf r(s'))$ it can be observed that $K(\mathbf x,\mathbf  x')$ is the path integral solution to the wick-rotated Schr{\"{o}}dinger equation, integrated over all ``times", but where time in this system is equal to spatial path length $s$. We have introduced the variable $s$ here, which has units of $[L]^2$. This variable characterizes distance along a path. We make the analogy more complete by characterizing the term in the exponential as a ``Lagrangian'' (as defined in equation ~\eqref{eq:GreenL}). We note that the origin of expression \eqref{eq:Green} needs to take care of Weyl ordering of operators, for a full discussion, see \cite{Gersten1987}.


\resub{This result isn't novel in the sense that it can be compared to the standard results obtained in linear response theory, however in this case we choose to represent the Green's function in path integral form, which is not standard in previous literature, for reasons of notation and interpretation. As shall be described immediately below, each of the terms that appear in the expression \ref{eq:Green} can be associated with a physical meaning that can make it more apparent how the presence of dissipative forces in a complicated system can affect the statistical weight of a particular state through understanding of the equilibrium landscape of $H$. This also allows us to bring certain physically justified approximations to bear in the evaluation of microstate probability. Moreover, and more practically, the usage of path integrals in quantum field theory has established methods for evaluation in different contexts, including anharmonic corrections, which we shall occupy ourselves with later in the manuscript, and poses quite general challenges for out of equilibrium systems.}

The derivation of this expression ~\eqref{eq:Green} results from mathematical considerations, interestingly however, we observe in the final expression ~\eqref{eq:Green} that the terms in the exponential have physical interpretation. The effective ``Lagrangian" (weight of the exponential) has three terms, which can be summarized as:
\begin{equation} \label{eq:totL}
    L = \text{total length of path}+\text{total power dissipated along path}+\text{total work done along path}
\end{equation}

This is using the fact that the total power dissipated along the path has been associated with the expression $\nabla H.\nabla H-\nabla^2 H$ in prior works \cite{Tome1997,Tome2006}. The total path length is given by $\big|\frac{\mathrm d \mathbf r(s)}{\mathrm d s}\big|^2$ and the work along the path is given by $\nabla H.\frac{\mathrm d \mathbf r(s)}{\mathrm d s}$. As the equilibrium system is conservative, the final term can be integrated out explicitly. This suggests that even in equilibrium, every two points in the state space are connected by the integral over all paths weighted by this quantity. Therefore, overly long paths have little effect, but so do paths where  a large amount of work is done by the system. Interestingly however, paths which also dissipate a lot of power also have minimal effects. This is a purely equilibrium function, and it could therefore be a source of confusion to speak of ``dissipation'' at equilibrium. While the average of  $\nabla H.\nabla H-\nabla^2 H$ can be shown to be exactly zero at equilibrium (meaning on average the system doesn't dissipate power), the same is not true for the evaluation of this quantity along a path connecting any two points.


The presence of this path dependent, non-local term makes this a difficult path integral to evaluate in general. However, a simplification arises in the cases that the Hamiltonian is slowly varying between points $\mathbf x$ and $\mathbf x'$, in this case $\frac{1}{4}\beta^2\nabla H(\mathbf r(s')).\nabla H(\mathbf r(s'))-\frac{1}{2}\beta\nabla^2 H(\mathbf r(s')) \approx 0$ we can ignore the ``difficult'' parts of the path integral, leading to a form:
\begin{equation} \label{eq:approx}
    G(\mathbf x, \mathbf x')\approx\frac{e^{\frac{\beta}{2}\left(H(\mathbf x)-H(\mathbf x')\right)}}{|\mathbf x-\mathbf x'|^{3N-2}}
\end{equation}
as was seen previously\cite{Osmanovic2021}. In most situations, as the effect of a perturbation at a point $\mathbf x$ declines very quickly in the total distance between two microstates $|\mathbf x-\mathbf x'|$, the effect of the dissipated power term will be rather small, \resub{as the variation in the Hamiltonian doesn't have an opportunity to be important before the other terms come to dominate the Green's function and therefore set $G$ to a infinitesimally small value.}

Therefore, one way of understanding the microscopic probability in a non-equilibrium situation involves summing all the paths (weighted by the Lagrangian given above) between those two points in equilibrium. This tells us the effect of a impulse of ``dissipated power" at a point $\mathbf x'$ on the microscopic probability at any point $\mathbf x$. 

\resub{For completeness, we note that the same function $G$ given above appears in the second order perturbation. The second order perturbation, which goes beyond linear changes in the nonconservative force $\mathbf F_{nc}$ is given by:
\begin{equation}
    d\log P_{2}(x)= \int\mathrm{d}\mathbf x' \int \mathrm d \mathbf x'' G(\mathbf x,\mathbf x')\left(\beta \mathbf F_{nc}(\mathbf x').\nabla'-\rho(\mathbf x')\right)G( \mathbf x',\mathbf x'')\rho(\mathbf x'')-\frac{1}{2}\left(\int \mathrm d \mathbf x' G(\mathbf x,\mathbf x') \rho(\mathbf F(\mathbf x'))\right)^2
\end{equation}
}
\resub{
In fact, using the operator given above the full microscopic probability for a system subject to a non-conservative force $\mathbf F_{nc}$ can be given in complete form as:
\begin{equation}
    P(\mathbf x)=\frac{1}{Z_{nc}}\exp\left(-\beta H(\mathbf x)+\log\left(\left[1-\int \mathrm d \mathbf x' G(\mathbf x,\mathbf x')\left(\beta \mathbf F_{nc}(\mathbf x').\nabla'-\beta^2 \nabla H(\mathbf x').\mathbf F_{\text{nc}}(\mathbf x')\right)\right]^{-1}\mu_0\right)\right)
\end{equation}
where $\mu_0$ is any constant and $Z_{nc}$ is a normalization constant. The term in the square brackets should be understood as an operator acting on $\mu_0$. This is assuming that the Neumann series given here converges. Therefore, given a closed form of $G$, we should be able to calculate arbitrary microscopic probability distributions under non-conservative forces. The next sections of the paper will be occupied with calculating integrals over $G$, and in which limits the simpler approximate forms presented here can be utilized.
}

\section{Evaluation of Green's function integrals}

In the previous section, we stated that the probability of a state of any system perturbed with a non-conservative force is mediated through its Green function, which is an equilibrium property of the system. A useful example to explore our expressions occurs when the Hamiltonian is harmonic:
\begin{equation} \label{eq:H}
    H_h(\mathbf x) = \frac{1}{2}\mathbf x.\underline \kappa.\mathbf x
\end{equation} 
Now we imagine a system with such a Hamiltonian being perturbed with a source term $\rho(\mathbf x)=\mathbf x.\underline B.\mathbf x $


A long calculation can follow where we can calculate what the first order effect of such a non-equilibrium perturbation would be. The details of this calculation are rather cumbersome, but are reproduced in the appendix \ref{app:appB}. The final result for the appearance of the first order log microscopic probability is given by:
\begin{equation}
    \log(P(\mathbf x)) \approx - \frac{1}{2} \beta \mathbf x.\underline \kappa.\mathbf x -\frac{1}{2} \beta^2 \mathbf x .  \underline C . \mathbf x
\end{equation} 
We have also introduced a new matrix $\underline C$, which results from integration over the Green's function, it is given by:
\begin{align} \label{eq:dissevo}
    &\int \mathrm d \mathbf x'  G(\mathbf x, \mathbf x') \left(\mathbf x'.\underline B.\mathbf x'\right) = \mathbf x.\underline C. \mathbf x +\text{Tr}(\underline \kappa^{-1}\underline B)\\
    &\underline C =  
        \int_0^\infty \mathrm d s  \exp(-\beta\underline \kappa s).\underline B.\exp(-\beta\underline \kappa s)
\end{align}
We can see that the Green's function integral over the term $\mathbf x'.\underline B.\mathbf x'$ reproduces a harmonic term, with an additional constant given by a matrix trace. In other words, a harmonic perturbation has a harmonic form after integration, plus some constant. This constant is irrelevant to the relative probability of any two states, however, \resub{and for divergence free non-conservative force the trace of these matrices will be equal to zero anyway.}

Of interest is how the integral in \eqref{eq:dissevo} applies to source functions that go as, for example $\rho(\mathbf x)= \Theta_{ijkl}x_ix_jx_kx_l$ where $\Theta$ is an object which gives coupling constants for entries $i,j,k,l$ of $\mathbf x$ (where repeated indices are summed, Einstein convention is employed, each index corresponds to one entry of the vector $\mathbf x$). This would correspond to, for example a system with an anharmonic Hamiltonian being driven by a non-conservative force. A similar calculation can be performed in this case (see appendix \ref{app:appB}), where the following result is obtained:
\begin{align}
    \int \mathrm d \mathbf x'  G(\mathbf x, \mathbf x') \left( \Theta_{ijkl}(x')_i(x')_j(x')_k(x')_l\right) =& \int_0^{\infty} \mathrm d s \Theta_{ijkl}(\chi(s))^i(\chi(s))^j(\chi(s))^k(\chi(s))^l \\&+\sum_{\Psi}\int_0^{\infty}\mathrm d s\mu(s)_{ij} \Theta_{ijkl}(\chi(s))_k(\chi(s))_l 
    \\&+ \sum_{\Psi}\mu(s)_{ij}\mu(s)_{kl}\Theta_{ijkl}
\end{align}
where the sum over $\Psi$ corresponds to sums over relevant permutations in the indices $i,j,k,l$ (the method of generation is given in the appendix) and we have introduced the matrix $\underline \mu(s)$ and vector $\boldsymbol \chi(s)$ that are given by:
\begin{align}
    \underline \mu(s)&= \left(\beta \underline \kappa)^{-1}.(\underline I - \exp(-2 \beta s \underline \kappa)\right)\\
    \boldsymbol \chi(s)&= \exp\left(-\beta \underline \kappa s\right).\mathbf x
\end{align}
$\exp$ here refers to the matrix exponential, and $I$ is the identity matrix. The derivations of these are also given in the appendix. 

From the above, it can be seen that given a base harmonic Hamiltonian, the Green's function can be applied mechanically onto any kind of source function $\rho(\mathbf x)$, reminiscint of Feynman rules. For example, to  source function going as $\sim x^6$, one could apply the following rule: contract the six index tensor with the matrix $\underline \mu$, zero, one, two and three times, and for any remaining indices contract them over the vector $\boldsymbol \chi$. The only remaining facet is to be careful over the indices being contracted.

We can combine the above discussion with the expression for the Green's function in a generic system, not just for a harmonic Hamiltonian (see appendix \ref{app:appC}). We imagine the equilibrium Hamiltonian is given by:
$H(\mathbf x)=H_{h}(\mathbf x) + V(\mathbf x)$ where $H_h$ are the parts of the Hamiltonian that go at most quadratically in $\mathbf x$.

In such a scenario, the Green's function is given by\cite{Wipf2013}:
\begin{equation} \label{eq:fullGreen}
    G(\mathbf x'',\mathbf x') =e^{\frac{\beta}{2}\left(V(\mathbf x'')-V(\mathbf x')\right) }\int_0^{\infty}\mathrm d s G_h(\mathbf x'',\mathbf x',s) \exp\left[\int_0^s \mathrm d s' V_a\left(\frac{\delta}{\delta \mathbf J(s')}\right)\right]\exp(-W[\mathbf J])\big|_{\mathbf J=0}
\end{equation}

Where the function $V_a$ is given by:
\begin{equation}
V_a(\mathbf x) = \nabla V(\mathbf x).\nabla V(\mathbf x) + 2 \nabla V(\mathbf x).\nabla H_h(\mathbf x) - \nabla^2 V(\mathbf x)
\end{equation}

, and $W[\mathbf J]$ is a functional of an arbitrary function $\mathbf J$ and is given by:

\begin{equation}
    W[J]=\int \mathrm d s \mathbf J(s).\mathbf x_{cl}(s) + \frac{1}{2}\int \mathrm d s\int \mathrm d s' \mathbf J(s).\underline G_D(s,s').\mathbf \underline \mathbf J(s')
\end{equation}
where $\underline G_D(s,s')$ is an analytic functions given in the appendix and $G_h(\mathbf x,\mathbf x',s)$ is given by:
\begin{equation}
G_h(\mathbf x,\mathbf x',s) = e^{\frac{\beta}{2}\left(H_h(\mathbf x'')-H_h(\mathbf x')\right)}\left(\frac{1}{2\pi}\right)^{D/2}\sqrt{\det\left(\frac{1}{2}\underline \kappa\text{csch}\left(\underline \kappa s\right)\right)}\exp\left(-S_{cl}\right)
\end{equation}
with $S_{cl}$ as the classical action integrated over the classical path, which can be found in the appendix.

The above introduced machinery appears formidable, but can be utilized in practice to derive corrections for anharmonic systems subject to non-conservative forces. As the applications of the Green's functions integrals are mechanical, all the above taken together can be used to derive leading order terms for anharmonic systems subjected to non-conservative forces. We shall see that interesting effects don't require the employment of the entirety of this machinery. Equation \eqref{eq:fullGreen} is exact, nevertheless it is difficult to use in practice, as is the usual story with anharmonic systems. To that end we investigate progressively more accurate representations of equation \eqref{eq:fullGreen}, firstly by ignoring the terms going as $V(\mathbf x)$ in the exponential, the harmonic approximation:
\begin{equation} \label{eq:harm}
G(\mathbf x'',\mathbf x')\approx \int_0^{\infty}\mathrm d s G_h(\mathbf x'',\mathbf x',s)
\end{equation}
then doing integrations over this kernel with the endpoint values (anharmonic integral approximation):
\begin{equation}\label{eq:anharmI}
G(\mathbf x'',\mathbf x')\approx e^{\frac{\beta}{2}\left(V(\mathbf x'')-V(\mathbf x')\right) }\int_0^{\infty}\mathrm d s G_h(\mathbf x'',\mathbf x',s)
\end{equation}
and finally, by including the first order effect of the dissipation along the path at equilibrium (anharmonic integral with dissipative correction approximation)::
\begin{equation} \label{eq:aIwithDC}
G(\mathbf x'',\mathbf x') \approx e^{\frac{\beta}{2}\left(V(\mathbf x'')-V(\mathbf x')\right) }\int_0^{\infty}\mathrm d s G_h(\mathbf x'',\mathbf x',s)\left(1+\int_0^s \mathrm d s' V_a\left(\frac{\delta}{\delta \mathbf J(s')}\right) \right)\exp(-W[\mathbf J])\bigg|_{\mathbf J=0}
\end{equation}

\subsection{Single Brownian particle in harmonic or anharmonic trap subject to active noise}

A useful application of the previous equations is in the calculation of the probability distribution of a particle bound in some anharmonic or harmonic potential subject to some active noise $A$. The harmonic form of this problem is a canonical model in prior literature\cite{Dabelow2019}. Such a system can be described by the following equations: 
\begin{align} \label{eq:sys}
    \gamma \frac{\mathrm d x}{\mathrm d t} &= - q^{(\alpha)}x^{\alpha} + a A(t)+  \xi(t) \\
    \label{eq:Ae}\gamma \frac{\mathrm d A(t)}{\mathrm d t} &= -\frac{\gamma}{\tau_a} A(t)  + \xi(t)
    \end{align}

where $\gamma$ is the viscous damping, the white noise $\xi(t)$ obeys the following relationship $\langle  \xi(t) \xi(t') \rangle = 2 k_b T \gamma \delta(t-t')$ with a temperature $k_b T$, $\tau_a$ is some timescale of relaxation of the active noise $A$, and $a$ characterizes the magnitude of the impact of that noise on the spatial variable $x$. The integer $\alpha$ is $\alpha=1$ for a harmonic system and $\alpha=3$ for an anharmonic one. The parameter $q^{(\alpha)}$ gives the magnitude of the restoring force. We include the superscript $(\alpha)$ to make clear that the units of $q^{(\alpha)}$ will vary depending on $\alpha$, $[q^{(\alpha)}]=[E][L]^{-\alpha+1}$. The conservative and non-conservative parts of the system \ref{eq:sys} can be found via symmetrizing the forces, yielding $\mathbf F_{c} = \left(- q^{(\alpha)}x^{\alpha},-\frac{\gamma}{\tau_a}A\right)$ and $\mathbf F_{nc}=\left(a A,0\right)$. Using the source function defined in equation \eqref{eq:source}, we have for system \eqref{eq:sys}
\begin{equation}
    \rho(\mathbf x) = q^{(\alpha)}a x^{\alpha}A
\end{equation}

We can go through the process of applying the machinery of the previous section to such a source term. We can also ask what level of approximations of the Green's function is necessary for the calculation of accurate results. For a harmonic system ($\alpha=1$), the matrix $\kappa $ quoted in the previous section is given by $\kappa=\begin{bmatrix}
    &q^{(1)},&0 \\ &0,&\gamma/\tau_a
\end{bmatrix}$, after performing the integral given in \eqref{eq:dissevo} we obtain for the first order perturbation:
\begin{equation}
    P_{\text{harmonic}}(x,A)\sim\exp\left(-\frac{\beta}{2}q^{(1)}x^2 -\frac{\beta}{2}\frac{\gamma}{\tau_a}A^2+\beta \frac{a q^{(1)} \tau_a}{q^{(1)} \tau_a+\gamma }x A\right)
\end{equation}
where $\beta=1/k_bT$. As expected, the introduction of a driving noise $A$ onto the spatial variable $x$ leads to a term in the probability that correlates $x$ and $A$. When viewed in the space of only $x$ (integrating out over $A$), this would correspond to increased variance of $x$, as would be expected and is known from an active noise acting on a particle.

A similar calculation (see appendix) can be performed for an active particle in an anharmonic trap ($\alpha=3$), yielding:
\begin{equation} \label{eq:anharmonicdist}
    P_{\text{anharmonic}}(x,A)\sim\exp\left(-\frac{\beta}{4}q^{(3)}x^4 -\frac{\beta}{2}\frac{\gamma}{\tau_a}A^2+c x A + d x^3 A\right)
\end{equation}
which is the first term for $q^{(3)}$ is small. Full integration over the anharmonic terms can be performed, however, a simple approximation results if we take only the leading order correction over the harmonic form of the Green's function (which we call the \textit{harmonic approximation} introduced in the previous section), which leads to the form of the variables $c$ and $d$, which goes as:
\begin{align} \label{eq:cform}
    c&=\frac{3 a q^{(3)} \tau _a^2}{\gamma ^2} +\mathcal{O}(q^2 \tau_a^4/\gamma^4) \\
    d&= \frac{\beta a q^{(3)} \tau _a}{2\gamma } +\mathcal{O}(q^2 \tau_a^3/\gamma^3)
\end{align}
Whereupon the rest of the series can be seen as including terms of higher powers of $\tau_a/\gamma$

\begin{figure}[!h]
    \begin{center}
    \includegraphics[width=90mm]{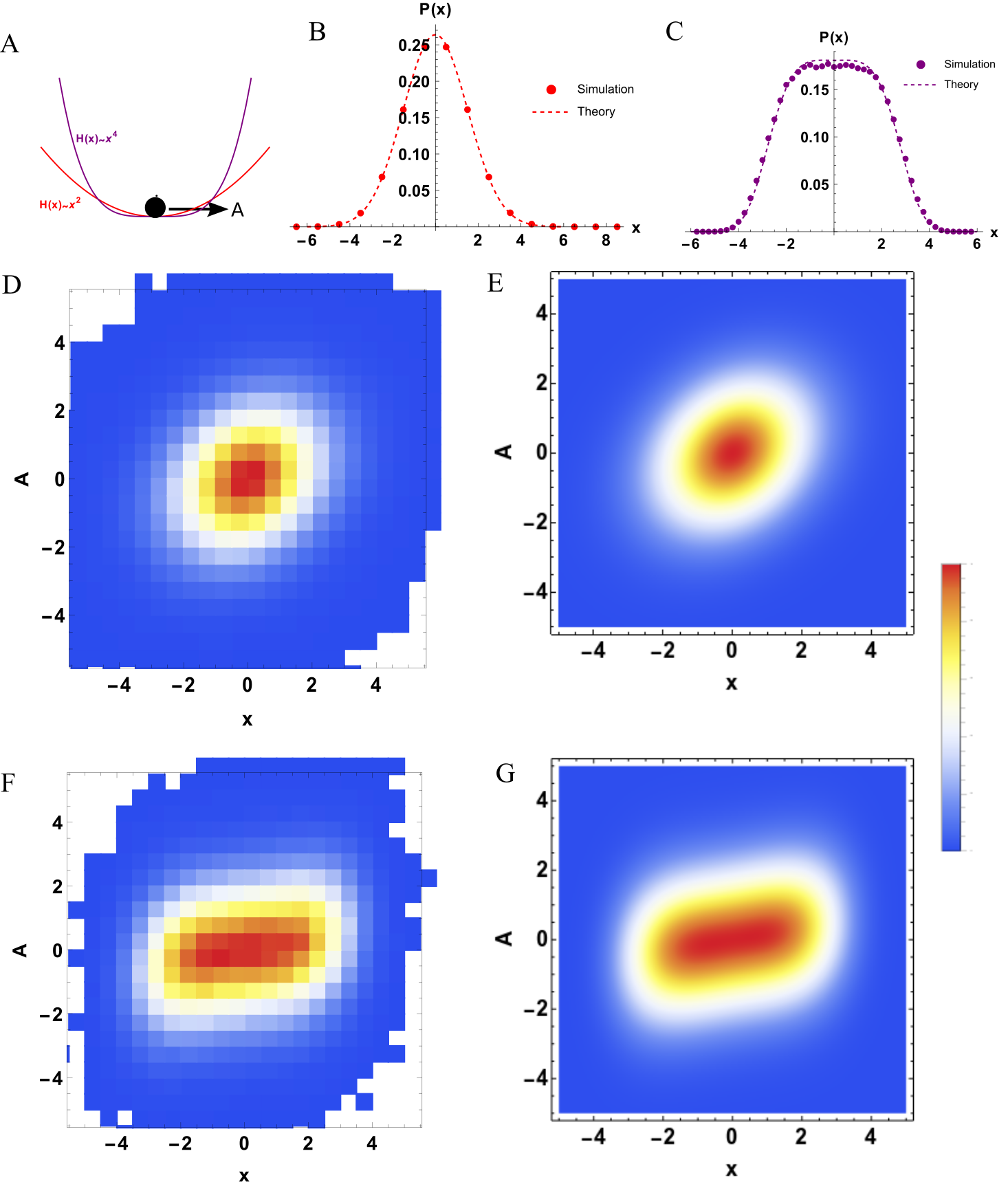}
    \caption{ The impact of an active noise $A$ on a Brownian particle in a harmonic or anharmonic well. A) schematic showing the particle in two different kinds of wells. B) Simulations of a Brownian particle in a harmonic well compared to theoretical predictions of the same arising from the first order perturbation for $q^{(1)}=1,\gamma=10,\tau_a=10,a=0.5$ C) The same as in B but for an anharmonic well with  $q^{(3)}=0.1,\gamma=10,\tau_a=10,a=0.5$ (taking the anharmonic integral approximation). D) and E) comparison of the full stationary probability in the space of $x$ and $A$ for the harmonic system with the parameters just mentioned, measured in simulation (E) and in theory (F). A coupling going as $xA$ is introduced by the non-conservative force. F) and G), the same but for the anharmonc system. As in the harmonic system, a coupling going as $xA$ is visible.   }
    \label{fig:fig1} %
    \end{center}
    \end{figure}

Comparison of analytic results with simulations of a Brownian particle under an active force are presented in figure \ref{fig:fig1}. For ease of comparison, we compare both harmonic and anharmonic trapping. It is plainly observable that a term of the form $x A$ is present in the probability density. Presented are also the probability densities integrated out over the active noise $A$ in figures \ref{fig:fig1}B and \ref{fig:fig1}C. The full microscopic probability densities are shown for both theory and simulation in D-G, and display good agreement.

 At first glance, it is somewhat surprising that we can obtain reasonable results in this regime, as we are perturbing around the harmonic system where there is no harmonic contribution to the variable $x$ at equilibrium, and $\gamma/\tau_a$ is not very large in comparison to $a$. This points to some robustness in the perturbative scheme developed thus far.  

\begin{figure}[!h]
    \begin{center}
    \includegraphics[width=90mm, frame]{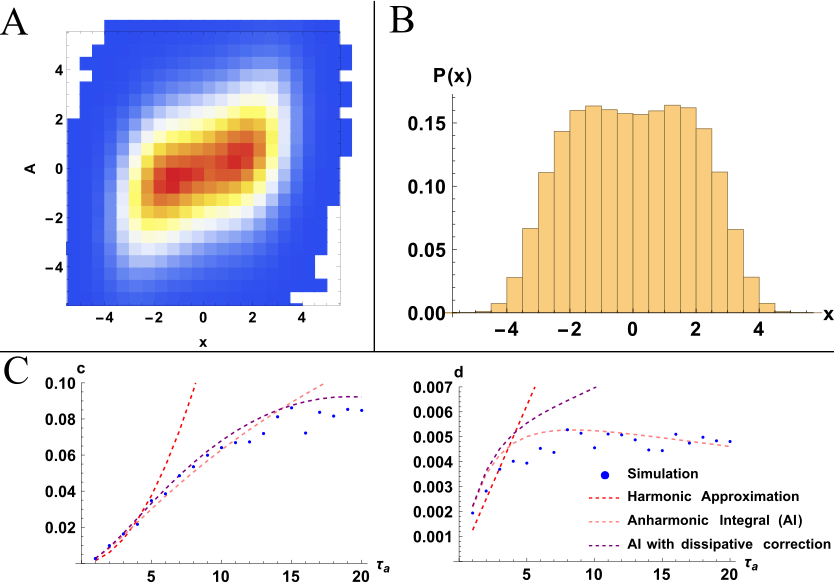}
    \caption{ Probability distributions under larger magnitudes of the active noise. A) the probability distribution in both $x$, $A$, from which a coupling going as $x^3 A$ is visible. B) Integration over $A$ yields the probability distribution of the particle in the trap. The maxima of this distribution is now away from $x=0$. C) Simulations of an active Brownian particle in an anharmonic trap where the measured distributions are fit to equation \ref{eq:anharmonicdist} for the parameters $c$ and $d$. The values of these parameters can also be calculated in theory under different approximations (dashed lines) where it can be seen that approximate forms of the Green's function nevertheless capture the correct scaling up to quite large $\tau_a$. }
    \label{fig:fig2} %
    \end{center}
    \end{figure}

One important qualitative result is the term going as $x A$, which appears in both harmonic and anharmonic traps, but we have yet to discuss its impact. Generally, the Green's function integrated out over a quartic source function will bring out a \textit{harmonic} contribution (i.e. $xA$). If we use the following integral identity: $\int \mathrm d A \exp(-c_1 x^4-c_2 A^2 + c_3 x A)=\text{constant}\times\exp\left(\frac{c_3^2 x^2}{4 c_2}-c_1 x^4\right)$ it can be observed that an anharmonic system subject to non-conservative forces will begin to accumulate probability away from $x=0$ (which is its equilibrium minima) as the maxima of the probability will now occur at $x=c_3/\sqrt{8 c_1c_2}$. Such an effect isn't observable for the magnitudes of non-conservative forces in figure \ref{fig:fig1}, but if the non-conservative force is made stronger, it can be seen; as in figure \ref{fig:fig2}B where probability begins to accumulate away from $x=0$, in contrast to the equilibrium system. Stronger drives also increase the importance of the term going as $x^3 A$, as can be observed in figure \ref{fig:fig2}A.

  What matters here is not so much the magnitude of the factor $c$, but the \text{form} of the coupling $x A$. This suggests less exact methods and/or approximate Green's funcions may get the precise magnitude of the coupling terms incorrect, but still obtain qualitatively correct couplings. More generally, Green's function integrals over \textit{any} anharmonic system should bring out terms of lower order whose magnitude can be important for larger amounts of non-conservative force. Therefore, accumulation away from the equilibrium mean should be expected as a generic result of anharmonic systems under non-conservative forces\cite{Das2018}, as contrasted to harmonic systems.

The approximate expressions given in equation \eqref{eq:cform} can be compared against simulations with strong drives, by increasing $\tau_a$. The resulting measured distributions can then be fit to the form $P(x)=\exp(-f_1(\beta/4)q^{(\alpha)} x^4-f_2\beta(\gamma/2\tau_a)A^2+ c x A + d x^3 A)$ and the values of $c$ and $d$ can be measured and compared to against expectations from the theory, which is seen in figures \ref{fig:fig2}C (we also include additional fitting parameters $f_1$ and $f_2$ to account for higher than first order effects in the simulations). The value of the parameters can be computed not only for the simple approximate expressions presented above, but for more detailed approximations to the Green's function $G(\mathbf x,\mathbf x')$. We focus on the two further approximations given previously, the one given in equation 
\eqref{eq:anharmI}  and the result for the first order perturbation of the dissipation-along-the-path contribution to the path integral arising in equation \eqref{eq:aIwithDC}. This allows us to probe the importance of different contributions to the Green's function in the magnitude of the coupling constants $c,d$. Figure \ref{fig:fig2}C shows that the approximation \eqref{eq:approx} is excellent even up to quite large values of $\tau_a$. This is notable as at large $\tau_a$ $(>10)$, the non-conservative force exceeds the conservative one. When the additional dissipation-along-the-path contribution is included, the value of $c$ measured in simulation is better captured, though the prediction for $d$ actually becomes worse. We mentioned previously that the contribution of the dissipation along the path for the Green's function becomes important only when the Hamiltonian varies quickly, which is why it's introduction is important for capturing the value of $c$ at large $\tau_a$, for under such noise the system is probing the quartic region of the Hamiltonian and therefore the terms going as $\nabla.H.\nabla H-\nabla^2 H$ become non-negligible.

As a final point, the above illustrates another far simpler method towards discovery of modifications of equilibrium distributions in the presence of non-conservative forces. By symmetry, the contributions to the probability distribution of the non-conservative force have odd parity $f(-x)=-f(x)$. Therefore, we can reason out the additional terms in a quartic system as by necessity being either $xA,x^3 A$ or $A^3 x$. Dimensional analysis can yield the approximate scalings of the prefactors. Such a method would not be able to derive numerical prefactors, however when attempting to construct effective theories these could simply be parameterized.

\section{Larger systems and mode space analysis: Application to Polymers}

Using some of the results from the previous section, we can briefly try to learn about the properties of larger systems based on the modifications of the microscopic probability that occurs due to non-conservative forces. Larger systems pose quite general challenges for microscopic descriptions as when the number of degrees of freedom is large, it is difficult to even represent microscopic probabilities. Thus, we want to describe some more macroscopic observables through consideration of the microscopic probability distributions derived in the previous section. A natural candidate system for larger systems are active polymers \cite{eisenstecken_2017,vandebroek_2015,winkler_2017,osmanovi_2017}, given that these have both non-conservative forces and harmonic (or anharmonic) backbone potentials. Such a system can be thought of as a collection of coupled particles in wells of the previous section. We imagine a polymer of beads each of whose monomers $i$ and $i+1$ with position vectors $\mathbf x_i$ and $\mathbf x_{i+1}$ are connected by a spring potential of the form:
\begin{equation}
    \phi(\mathbf x_{i+1},\mathbf x_i) = \frac{1}{\alpha+1}q^{(\alpha)} (\mathbf x_{i+1}-\mathbf x_i)^{\alpha+1}
\end{equation}
where once more $\alpha=1$ is a harmonic system and $\alpha=3$ is an anharmonic one. The full Hamiltonian of this polymer can be written as $H=\sum_i^N \phi(\mathbf x_{i+1},\mathbf x_i)$. We subject each monomer in this system to an active noise of the form given in the previous section, such that the time evolution equation of each individual monomer will be given by:
\begin{equation}
    \gamma \frac{\mathrm d \mathbf x_i}{\mathrm d t} =  -\nabla_{\mathbf x_i} \left(\phi(\mathbf x_{i+1},\mathbf x_i)+)\phi(\mathbf x_{i},\mathbf x_{i-1})\right) + a \mathbf A_i(t) + \boldsymbol \xi(t)
\end{equation}
where each entry of vector $\mathbf A_i$ obeys equation \eqref{eq:Ae}. This is an extended version of the system in the previous example. We may now ask about the stationary properties of such a system, but it would be useful to compare the results to a similar equilibrium system. We already saw that the tendency of the addition of active noise is to bring forth a harmonic term for an anharmonic Hamiltonian that leads to accumulation away from the equilibrium minima. We therefore compare these driven harmonic and anharmonic polymers with an equilibrium polymer with the following potential:
\begin{equation} \label{eq:equipolymer}
\Phi(\mathbf x_{i+1},\mathbf x i) = -\frac{1}{2}k(\mathbf x_{i+1}-\mathbf x_i)^2 +\frac{1}{4}q (\mathbf x_{i+1}-\mathbf x_i)^4
\end{equation}
where $k,q$ are positive constants. We concentrate on two different aspects of the polymer. One particular macroscopic quantity which is natural to study in polymers is the average end to end distance $\langle R^2\rangle$ which is the average distance between the two endpoints of the polymer. Another quantity we wish to analyze is how the modes of the polymer are modified in the presence of drive. Mode space analysis is standard in polymer systems, defining a set of collective coordinates of the entire polymer, i.e. $\tilde{\mathbf x}=\underline Q .\mathbf x$ where $\underline Q$ is a matrix of eigenvectors. Each mode is an independent degree of freedom with some associated eigenvalue $\lambda$. We know that in equilibrium that each of these modes should follow the distribution $\sim \exp(-\beta \lambda_i \tilde x^2 )$. In other words, in a harmonic system under equilibrium, each mode will have a variance that goes as the reciprocal of its eigenvalue. By analyzing the fluctuations of modes in an active system we can see roughly how ``unusual" such a system is compared to an equilibrium one if we divide through by $\exp(-\beta \lambda_i \tilde x^2 )$. By equipartition, for an equilibrium system, the variance of each mode divided by $1/\lambda_i$ should be a constant, however this no longer holds for non-equilibrium systems. Moreover, anharmonic polymers can be linearized around their minima, yielding similar mode space distributions as harmonic polymers. The analysis of all these quantities is presented in figure \ref{fig:fig3}.

\begin{figure}[!t]
    \begin{center}
    \includegraphics[width=180mm, frame]{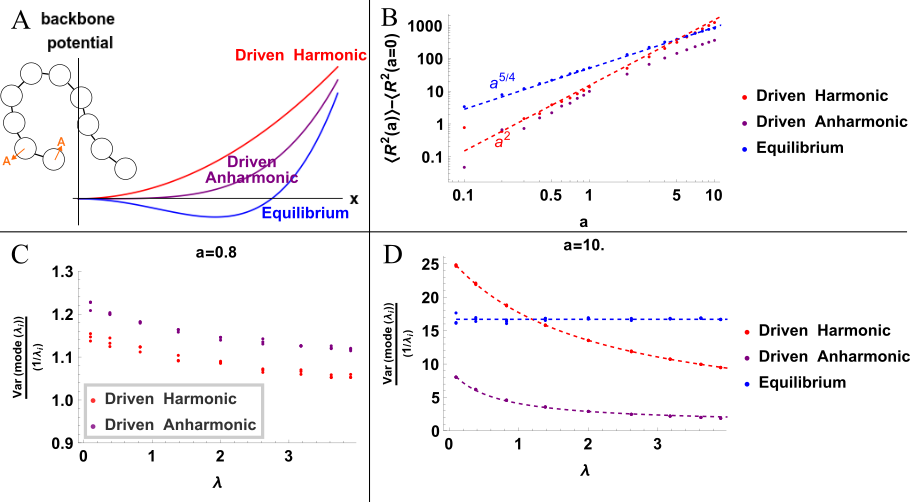}
    \caption{ Simulations of an active polymer. A) The polymer is being driven by some active noise $A$, with $\tau_a=10.$,$\gamma=10$,$k_b T=2.$, and we test our results for both a harmonic and anharmonic backbone potential, as well as against an equilibrium (no active noise) polymer containing both harmonic and anharmonic backbone potentials (equation \eqref{eq:equipolymer}) (see text for reasoning). B) Measured end to end distance for harmonic ($q^{(1)}=1$) and anharmonic($q^{(3)}=0.1$) polymers under active noise. Also shown, an equilibrium polymer with $k=a$ and ($q^{(3)}=0.1$). The size of the active harmonic polymer scales with noise as $a^2$, and the equilibrium polymer with $k^{5/4}$. The anharmonic active polymer will scale with a crossover between these two regimes due to the interplay of backbone potential and active noise. C) Variances of modes of the polymer measured against their eigenvalues. In the absence of active noise, these lines would be flat, however, when active noise is introduced the modes with smaller eigenvalues are excessively affected. D) Same as C) but at higher values of the drive.  }
    \label{fig:activepolymer}
    \end{center}
    \end{figure}

In figure \ref{fig:activepolymer}B we can observe several features that arise directly from the consideration of our microscopic model in the previous section. Firstly, the interaction between two neighboring monomers being driven by an active force $A$ of magnitude $a$ should have a microscopic distribution that has the form $P(x,A)\sim\exp(-c_1 x^4-c_2 A^2+ a c_3 x A)$ which leads to an effective microscopic probability arising between the monomers of form $P(x)\sim \exp({\frac{a^2 c_3^2 x^2}{4 c_2}-c_1 x^4})$. This means that for low $a$, the effective interaction between monomers is unmodified, but as $a$ is increased to $a>1$, the monomers should start to accumulate at a preferred length given by the interplay of the potential and the active noise. This can be compared to the equilibrium model defined by equation \eqref{eq:equipolymer}. In such a model, the total end to end distance scales against the parameter $k$ as $k^{5/4}$, as can be seen in figure \ref{fig:activepolymer}B. This is the same scaling observed in the active model against $a$ when $a>1$, indicating that the accumulative effect of the active noise has a similar effect as the interaction in the equilibrium model. We may contrast this with the active harmonic model, where the end to end distance  of the polymer chain always swells as $a^2$, as there is no accumulative effect away from the equilibrium minimum in harmonic models. 

Differences to the equilibrium model can be observed when looking at the variances of the modes of the polymer. It can be shown (see appendix \ref{app:appD}) how each mode is affected by the presence of the non-conservative force. As the modes are orthogonal degrees of freedom, in an equilibrium system we should expect by equipartition that the total energy per mode is the same, as is observed in simulations of the equilibrium model (and can be seen in figure \ref{fig:activepolymer}D). In contrast, for active systems equipartition no longer holds. What can be derived in the harmonic system and also holds in the anharmonic system is that modes with smaller eigenvalues (i.e. modes that have greater fluctuations under equilibrium) are more affected by the drive relative to their equilibrium values than the modes with large eigenvalues. For an active polymer, this means that the fluctuations of each mode should follow a form given by approximately $1/(1+\lambda^2)$ (see appendix \ref{app:appD}) where $\lambda$ is the eigenvalue of the mode.  As anharmonic systems also adopt harmonic distributions around the minima in the free energy, it is observed empirically (in both figure \ref{fig:activepolymer}C and \ref{fig:activepolymer}D). This effect becomes more pronounced the stronger the active noise $a$ is.

\section{Discussion and Conclusion}

In the preceding, we have demonstrated that the microscopic probability for a system perturbed by a deterministic non-conservative force or due to coupling with different heat baths can be shown to be equal to the equilibrium system with an additional term appearing as an integral over a microscopic stationary response (or Green's) function. This response function is the same for both of the above non-equilibrium perturbations. The microscopic response function is a path integral in the unperturbed system. In equilibrium, any two states $\mathbf x$ and $\mathbf x'$ are connected by the sum of all paths between them weighted by the path length, work and dissipated power associated with each of the paths. The modification of the microscopic probability is analytically solvable for harmonic systems. This formulation can be applied, though not without some difficulty, to anharmonic systems as well. It can be shown that purely anharmonic systems should see harmonic terms arising in their microscopic probability, from which it is apparent that accumulation of probability away from their equilibrium minima is a generic feature of anharmonic but not harmonic systems. Finally, we used the results we obtained for a Brownian particle in an anharmonic trap to understand the properties of anharmonic polymers under active noise.

The microscopic stationary response function was constructed through consideration of the dynamical processes involved, and yet its final form has a more physical interpretation. It would be interesting to start from the effective Lagrangian formulation as it appears in ~\eqref{eq:totL} and apply it to a completely different system without considering any dynamics and observe whether it can recapitulate the stationary distribution successfully. This would be suggestive as to the importance of the response function we have shown in this paper. 

The path integral that represents the response function is summed over all path lengths, where the path length takes a form that looks suspiciously like time. In this paper, we have ignored dynamic effects, this can prompt us to ask the extent to which the effect of time appears as the path length representation that is already present in the over-damped system. Following on from this, we have ignored inertial effects here, but the theory could be modified to account for them. One could then construct dynamical response functions with explicit time, and test for modifications to the fluctuation-dissipation relation.\cite{Caprini2021,Caprini_2021}

The fact that the soft modes (modes with the smallest eigenvalues) of the equilibrium system are most affected by the presence of non-equilibrium driving is intriguing. Interesting phenomena due to dissipative forces would arise in the case where the eigenspectrum of the original system was quite broad. In this case, some modes are basically unaffected by the drive, however the modes which were most strongly fluctuating in the equilibrium system are those which will be most strongly affected by the dissipative forces. This could lead to potentially quite exotic states due to the modification of the microscopic probabilities.


In terms of connection with experimental systems, active matter systems can be mapped onto this framework when the proper degrees of freedom the system are considered\cite{Osmanovic2021}, to this we would add experiments that involve heating different parts of a system to different temperatures \cite{Grosberg2015}. There has been recent interest in active matter with what has been deemed odd viscosity\cite{Banerjee2017,Markovich2021}, as these systems also have non-conservative forces they would be a natural candidate for investigating how the microscopic phenomena we have discussed here would bridge length scales to more macroscopic quantities. 

\bibliography{mypaper.bib}

\appendix

\section{Microscopic probability of overdamped system with multiple temperature baths} \label{app:appA}
The stationary state of the Smoluchowski equation for the log probability $\phi$ of a system with temperature $k_b T$ is given as the solution to\cite{risken_1989}:
\begin{equation}
    \nabla.\left(\exp(-\phi)\left(-\mathbf F - k_b T \nabla \phi \right)\right)=0
\end{equation}
in order to account for different temperatures, we can modify the equation:
\begin{equation}
    \nabla.\left(\exp(-\phi)\left(-\mathbf F - \underline \beta^{-1}. \nabla \phi \right)\right)=0
\end{equation}

where $\underline\beta^{-1}$ is a diagonal matrix where each entry on the diagonal is given by a value $k_b T_i$ corresponding to the temperature of that degree of freedom. This equation can be manipulated to give:
\begin{equation}
    \nabla.\left(\exp(-\phi)\underline \beta^{-1} \left(- \underline \beta.\mathbf F -\nabla \phi \right)\right)=0
\end{equation}
Splitting the vector field $ \underline \beta.\mathbf F  = \mathbf f^{c}+ \mathbf  f^{a}$ where $\mathbf f_c =  -\nabla \phi$ is a solution if $\nabla.\left(e^{-\phi}\underline\beta \mathbf f^{a}\right)=0$. Assuming that $\mathbf F=-\nabla H_0 $ this leads to the following equation:
\begin{equation}
    -\nabla H . \nabla \phi + \nabla \phi. \underline \beta^{-1} . \nabla \phi + \nabla^2 H -\nabla .(\underline \beta^{-1}.\nabla \phi)=0
\end{equation}
where if every entry into the matrix $\beta$ is equivalent, this reproduces the Boltzmann distribution. 

We can write the temperature matrix as a modification around a base temperature $\beta_1^{-1}$:
\begin{equation}
    \underline \beta^{-1} = \beta_1^{-1} \underline I + \beta_2^{-1} \underline D
\end{equation}
we redefine $H$ as including the base temperature $\beta_1^{-1}$
\begin{equation}
    H\coloneqq \beta_1 H_0
\end{equation}
We therefore have the following equation in $\phi$:
\begin{equation}
    -\nabla H.\nabla \phi +|\nabla \phi|^2+\alpha \nabla \phi.\underline D.\nabla \phi +\nabla^2 H -
    \nabla^2 \phi - \alpha \nabla.(\underline D. \nabla \phi)
\end{equation}
where $\alpha = \beta_1 \beta_2^{-1}$.

A perturbative expansion in $\alpha$, $\phi=\sum_{n=0}^{\infty} \alpha^n \phi_{(n)}$ gives for the term $\phi_{(k)}$:

\begin{equation} \label{eq:tP}
    -\nabla H.\nabla  \phi_{(k)}+\sum_{|n+m|=k}\nabla  \phi_{(n)}.\nabla  \phi_{(m)}+\sum_{|n+m+1|=k}\nabla  \phi_{(n)}.\underline D.\nabla  \phi_{(m)}  -\nabla^2  \phi_{(k)} - \nabla.(\underline D.\nabla \phi_{((k-1)})=0
\end{equation}
for $k>0$ and where $\phi_{(0)} = H$
which leads to the first order perturbation:
\begin{equation}
(\nabla^2-\nabla H.\nabla) \phi_{(1)} = \nabla.(\underline D.\nabla H) + \nabla  H.\underline D.\nabla  H
\end{equation}
It can be seen through repeated application of \ref{eq:tP} that the Green's function is of the form specified in the main text for every perturbation. 

\section{Evaluation of harmonic path integral for non-conservative system} \label{app:appB}

We use the generic solution to a path integral with a harmonic Lagrangian for a system with a Hamiltonain given by $H_0(\mathbf x)=\frac{1}{2}\mathbf x. \underline \kappa.\mathbf x$, which is given through the following formula\cite{Grosche1995}:
\begin{equation}
    \int_{\mathbf r(0)=\mathbf x'}^{\mathbf r(t)=\mathbf x''} \!\!\!\!\!\!\mathcal{D}\left[\mathbf r(t)\right]e^{\frac{i}{\hbar}\int_0^t \mathrm d t'  L(\dot x,x)} = \left(\frac{1}{2 \pi i \hbar}\right)^{D/2}\sqrt{\det\left(-\frac{\partial^2 S_{cl}[\mathbf x'', \mathbf x']}{\partial x_a'' \partial x_b' }\right)}\exp\left(\frac{i}{\hbar} S_{cl}[\mathbf x'', \mathbf x'] \right )
\end{equation}

where we perform the wick rotation $\hbar \to -i $ (note the lack of Boltzmann factor here, it is in fact not introduced into this analysis due to the fictitious nature of the time variable we have). Our Lagrangian is given by:
\begin{equation}
L=\frac{1}{4}\dot{\mathbf x}^2 + \frac{1}{4} \mathbf x.\underline \kappa. \underline \kappa.\mathbf x - \frac{1}{2} \text{Tr}(\underline \kappa)
\end{equation}

The quantity $S_{cl}$ is the classical action evaluation along the classical solution $\mathbf x_{cl}$ satisfying the boundary conditions $\mathbf x_{cl}(0)=\mathbf x'$ and $\mathbf x_{cl}(s)=\mathbf x''$

The Lagrangian leads to the following Euler Lagrange equation:
\begin{equation}
    \frac{1}{2}\ddot{\mathbf x} = \frac{1}{2} \underline \kappa.\underline \kappa.\mathbf x
\end{equation}
which has the generic solution:
\begin{equation}
    \mathbf x_{cl}(\tau) =  e^{\underline \kappa \tau}\mathbf c_1 + e^{-\underline \kappa \tau}\mathbf c_2
\end{equation}
where we are free to choose the value of $v_0$ to satisfy the boundary conditions. This can be done so, yielding the following classical path:


\begin{equation} \label{eq:classicalpath}
    \mathbf x_{cl}(\tau) = \text{csch}(\underline \kappa s ) (\mathbf x'' \sinh (\underline \kappa \tau)+\mathbf x' \sinh (\underline  \kappa (s-\tau )))
\end{equation}

The evaluation of the classical action evaluated along this path can be performed (using the fact that all the hyperbolic functions are power series in $\underline \kappa$, and thus commute):
\begin{equation}
    S_{cl}=\frac{1}{4} \mathbf x' .\underline \kappa \text{coth}\left(\underline \kappa s\right). \mathbf x'+\frac{1}{4} \mathbf x'' .\underline \kappa \text{coth}\left(\underline \kappa s\right) .\mathbf x''-\frac{1}{2}  \mathbf x'' .\underline \kappa\text{csch}\left(\underline \kappa s\right). \mathbf x'-\frac{1}{2}\text{Tr}(\underline \kappa) s
\end{equation}
this allows us also the evaluate the Von Vleck determinant, which is given by:
\begin{equation}
    \sqrt{\det\left(-\frac{\partial^2 S_{cl}[\mathbf x'', \mathbf x']}{\partial x_a'' \partial x_b' }\right)}=\sqrt{\det\left(\frac{1}{2}\underline \kappa\text{csch}\left(\underline \kappa s\right)\right)}
\end{equation}

Finally, we can write the full expression for $K(\mathbf x'',\mathbf x')$:
\begin{align}
    K(\mathbf x'',\mathbf x') &= \int_0^{\infty}\mathrm d s\left(\frac{1}{2\pi}\right)^{D/2}\sqrt{\det\left(\frac{1}{2}\underline \kappa\text{csch}\left(\underline \kappa s\right)\right)}\exp\left(-S_{cl}\right) \\
    K(\mathbf x'',\mathbf x') &= \int_0^{\infty}\mathrm d s\left(\text{PFS}(s)\exp\left(-S_{cl}\right)\right)
\end{align}
where $\text{PFS}(s)=\left(\frac{1}{2\pi}\right)^{D/2}\sqrt{\det\left(\frac{1}{2}\underline \kappa\text{csch}\left(\underline \kappa s\right)\right)}$
Furthermore, the prefactors can be added leading to the full Green's function:
\begin{equation}
    G(\mathbf x'',\mathbf x') = \int_0^{\infty}\mathrm d s \text{PFS}(s)\times\exp\left(-S_{cl} +\frac{1}{4}\mathbf x''.\underline \kappa.\mathbf x''- \frac{1}{4}\mathbf x'.\underline \kappa.\mathbf x'\right)
\end{equation}

As an example, we can use standard Gaussian integrals to find the perturbation associated with a Harmonic source term (such as would arise from the action of different temperature baths) $\left(\mathbf x' .\underline B.\mathbf x'\right)$. We denote the difference of the log probability from equilibrium as $\delta(\log P )$. We have to calculate the following integral:

\begin{equation} \label{eq:int}
    \delta(\log P ) \sim \int_0^{\infty}\mathrm d s \text{PFS}(s)\int \mathrm d \mathbf x' \exp\left( -\frac{1}{2} \mathbf x'. (\underline M+ \frac{1}{2}\underline \kappa).\mathbf x'-\frac{1}{2} \mathbf x''. (\underline M- \frac{1}{2}\underline \kappa).\mathbf x'' + \mathbf b.\mathbf x' \right)\left(\mathbf x' .\underline B.\mathbf x'\right)
\end{equation}
where 
\begin{align}
\underline M(s) &= \frac{1}{2}\underline \kappa \text{coth}\left(\underline \kappa s\right) \\
\mathbf b(s)&= \mathbf{x''} \frac{1}{2} \underline \kappa\text{csch}\left(\underline \kappa s\right)
\end{align}
By completing the square, we can obtain the following equivalent expression for the exponential:
\begin{align}
    \delta(\log P ) \sim \int_0^{\infty}\mathrm d s\text{PFS}(s) \int \mathrm d \mathbf x'\exp\bigg(-&\frac{1}{2}\left(\mathbf x'-(\underline M(s)+\frac{1}{2}\underline \kappa)^{-1}\mathbf b(s))\right).(\underline M(s)+\frac{1}{2}\underline \kappa).\left(\mathbf x'-(\underline M(s)+\frac{1}{2}\underline \kappa)^{-1}\mathbf b(s))\right) \\ \nonumber &+ \frac{1}{2}\mathbf b(s).(M(s)+\frac{1}{2}\kappa)^{-1}.\mathbf b(s)-\frac{1}{2} \mathbf x''. (\underline M(s)-\frac{1}{2} \underline \kappa).\mathbf x''\bigg)
\end{align}
The substitution $\mathbf y(s)=\mathbf x'-(\underline M(s)+\frac{1}{2}\underline \kappa)^{-1}\mathbf b(s)$ allows us to compute the integral in equation \ref{eq:int}. This leads to the following expression:
\begin{align} \label{eq:FUG}
    \delta(\log P )  \sim \int_0^{\infty}\mathrm d s \text{PFS}(s)\int \mathrm d \mathbf y(s) &\exp\left(-\frac{1}{2} \mathbf y(s).(\underline M(s)+\frac{1}{2}\underline \kappa).\mathbf y(s)+ \frac{1}{2}\mathbf b(s).(\underline M(s)+\frac{1}{2}\underline \kappa)^{-1}.\mathbf b-\frac{1}{2} \mathbf x''. (\underline M(s)- \frac{1}{2}\underline \kappa).\mathbf x''\right)
    \\ \nonumber &\times \left(\mathbf y(s).\underline B.\mathbf y(s) + \mathbf b(s) .(\underline M(s)+\frac{1}{2}\underline \kappa)^{-1}.\underline B .(\underline M(s)+\frac{1}{2}\underline \kappa)^{-1}\mathbf b)  \right)
\end{align}
Where we have suppressed the terms linearly dependent on $\mathbf y$ outside the exponential, as these will integrate out to zero. This integral can be performed over $\mathbf y$ by using the standard properties of Gaussian integrals, leading to the following expression:
\begin{align} \label{eq:totI}
    &\sim \int_0^{\infty}\mathrm d s \text{PFS}(s) \left(\frac{(2\pi)^N}{\det(\underline M(s)+\frac{1}{2}\underline \kappa)}\right)^{1/2}\left(\text{Tr}\left((\underline M(s)+\frac{1}{2}\underline \kappa)^{-1}.\underline B \right)+ \mathbf b(s) .(\underline M(s)+\frac{1}{2}\underline \kappa)^{-1}.\underline B .(\underline M(s)+\frac{1}{2}\underline \kappa)^{-1}\mathbf b(s))\right) \\ \nonumber &\times \exp\left(\frac{1}{2}\mathbf b(s).(\underline M(s)+\frac{1}{2}\underline \kappa)^{-1}.\mathbf b(s)-\frac{1}{2} \mathbf x''. (\underline M(s)- \frac{1}{2}\underline \kappa).\mathbf x''\right)
\end{align}

This expression is then multiplied by the von Vleck determinant and other prefactors whose dependence we suppresssd earlier to lead to the full resultant expression for a Harmonic perturbation, once the integral over all times is performed. Though this expression appears difficult, remarkable simplifications occur. After all the experssions are reinserted, the term in the exponent is actually equal to zero.

The von vleck determinant $\text{PFS}(s)$ multiplied by the determinant arising from the integral is given by:
\begin{equation}
    \left(\frac{1}{2\pi}\right)^{N/2}\sqrt{\det\left(\frac{1}{2}\underline \kappa\text{csch}\left(\underline \kappa s\right)\right)}*\left(\frac{(2\pi)^N}{\det(\underline M+\frac{1}{2}\underline \kappa)}\right)^{1/2}
\end{equation}

The hyperbolic expressions can again be simplified, leading to:
\begin{equation}
    \left(\frac{1}{2\pi}\right)^{N/2}\sqrt{\det\left(\frac{1}{2}\underline \kappa\text{csch}\left(\underline \kappa s\right)\right)}*\left(\frac{(2\pi)^N}{\det(\underline M+\frac{1}{2}\underline \kappa)}\right)^{1/2}=\sqrt{\det\left(\exp(-\underline \kappa s)\right)} = \exp(-\text{Tr}(\underline \kappa)s/2)
\end{equation}
Where we use the property that the determinant of the matrix exponential being equal to the exponential of the trace. We can observe that this term cancels out with the term in the Lagrangian proportional to $\text{Tr}(\underline \kappa)$

Further simplifications exist:
\begin{equation}
    \mathbf b .(\underline M+\frac{1}{2}\underline \kappa)^{-1}.\underline B .(\underline M+\frac{1}{2}\underline \kappa)^{-1}\mathbf b) = \mathbf x'' \exp(-\underline \kappa s)\underline B \exp(-\underline \kappa s) \mathbf x''
\end{equation}

The term $\text{Tr}\left((\underline M(s)+\frac{1}{2}\underline \kappa)^{-1}.\underline B \right)$ appearing in equation \ref{eq:totI} doesn't have any dependence on $\mathbf x''$. In other words, it would change the probability uniformly up or down, however, this dependence disappears when we divide through by the partition function. It is important to bear this term in mind for higher order corrections, however.

We therefore have the following expression:
\begin{equation}
    \int_0^\infty \mathrm d s \mathbf x'' \exp(-\underline \kappa s)\underline B \exp(-\underline \kappa s) \mathbf x''
\end{equation}
for the modification in the probability. As seen in the main text.

\subsection{Integration over non-harmonic sources }

The preceding discussion can be applied to source functions of anharmonic form fairly straightforwardly. If we use the facts that we saw in the preceding discussion, that

\begin{equation}
    (\underline M(s)+\frac{1}{2}\underline \kappa)^{-1}.\mathbf b(s)    = \exp(-\underline A s)\mathbf x'' \equiv \boldsymbol \chi(s)
\end{equation}
\begin{equation}
    \left(\frac{1}{2}\mathbf b(s).(\underline M(s)+\frac{1}{2}\underline \kappa)^{-1}.\mathbf b(s)-\frac{1}{2} \mathbf x''. (\underline M(s)- \frac{1}{2}\underline \kappa).\mathbf x''\right) = 0
\end{equation}
\begin{equation}
    (\underline M(s)+\frac{1}{2}\underline \kappa)^{-1}= \left(\beta \underline \kappa)^{-1}.(\underline I - \exp(-2 \beta s \underline \kappa)\right) \equiv \underline \mu(s)
\end{equation}

We can take an arbitrary function of $\mathbf x'$ at equation \ref{eq:FUG} $f(\mathbf x')$, giving:
\begin{equation}
    \delta(\log P )  \sim \int_0^{\infty}\mathrm d s \text{PFS}(s)\int \mathrm d \mathbf y(s) \exp\left(-\frac{1}{2} \mathbf y(s).\underline \mu(s)^{-1}.\mathbf y(s)\right) f\left(\mathbf y(s)+\exp(-\underline A s)\mathbf x''\right)
\end{equation}

where such formulas can be evaluated using the identity\cite{zee2010quantum}:

\begin{equation}
    \int \mathrm d \mathbf y(s) \exp\left(-\frac{1}{2} \mathbf y(s).\underline \mu(s)^{-1}.\mathbf y(s)\right) f\left(\mathbf y(s)+\boldsymbol \chi(s)\right)=\left(\sqrt{\frac{(2\pi)}{\det(\underline \mu(s)^{-1})}}\right)^n 
     \nonumber \times \exp\left(\frac{1}{2} \mu(s)_{i,j}\frac{\partial}{\partial y_i}\frac{\partial}{\partial y_j}\right)f\left(\mathbf y(s)+\boldsymbol \chi(s)\right)\bigg|_{\mathbf y=0}
\end{equation}
for a generic function $f(\mathbf y)$ equal to some power series in $\mathbf y$, the above expression can be expanded to yield the sets of indices necessary for the summations in the main text.

\section{Anharmonic path integral for non-conservative system} \label{app:appC}

In order to calculate the effect of arbitrary perturbations, we must include the effect of anharmonicity, much of this discussion bears similarity with similar problems arising in quantum field theory\cite{zee2010quantum}, we recall that the definition of the Green's function is given by:

\begin{align}G(\mathbf x'',\mathbf x')&=\int_0^{\infty}\mathrm d s \int_{\mathbf r(0)=\mathbf x'}^{\mathbf r(s)=\mathbf x''} \!\!\!\!\!\!\mathcal{D}\left[\mathbf r(s)\right]e^{-\int_0^s \mathrm d s'  L(\dot{x},x)}
\\
&=e^{\frac{\beta}{2}\left(H(x)-H(x')\right) }\int_0^{\infty}\mathrm d s \int_{\mathbf r(0)=\mathbf x'}^{\mathbf r(s)=\mathbf x''}\!\!\!\!\!\!\mathcal{D}\left[\mathbf r(s)\right]e^{-\int_0^s \mathrm d s'  L(\dot{x},x)}
\end{align}

Our difficulty lies in evaluation of the path integral in the presence of anharmonic conservative forces in $L$. One way beyond this difficult is splitting the action into a baseline contribution which we can evaluate, and another contribution which we cannot.  $\int_0^s \mathrm d s'  L(\dot{x},x) = S_h+S_a$. We choose the baseline to be the harmonic path integral which we have just evaluated in the previous section. Then, conducting a Taylor series in $S_a$ we obtain:
\begin{equation} \label{eq:per}
   G(\mathbf x'',\mathbf x') = e^{\frac{\beta}{2}\left(H(x)-H(x')\right) }\int_0^{\infty}\mathrm d s \sum_{n=0}^{\infty}\frac{1}{n!}\int_{\mathbf r(0)=\mathbf x'}^{\mathbf r(s)=\mathbf x''}\!\!\!\!\!\!\mathcal{D}\left[\mathbf r(s)\right]e^{-S_h}(-S_a)^n
\end{equation}

In order to evaluate the path integrals in eq. \ref{eq:per} we introduce the following Lagrangian:
\begin{equation} \label{eq:lagso}
    L_J=\int_0^s \mathrm d s' \frac{1}{4}\dot{\mathbf x(s')}^2 + \frac{1}{4} \mathbf x(s').\underline \kappa. \underline \kappa.\mathbf x(s') - \frac{1}{2} \text{Tr}(\underline \kappa) + \mathbf J(s').\mathbf x(s') 
\end{equation}
By taking repeated functional integrals with respect to $J$, we can reproduce the terms appearing in eq. \ref{eq:per}. Crucially, with the Lagrangian given in \ref{eq:lagso}, the path integral can be evaluated analytically, giving:
\begin{equation}
    K(\mathbf x'',\mathbf x',s) = \int_{\mathbf r(0)=\mathbf x'}^{\mathbf r(s)=\mathbf x''}\!\!\!\!\!\!\mathcal{D}\left[\mathbf r(s)\right]e^{-\int_0^s \mathrm d s'  L_J(\dot{x},x)} =K_h(\mathbf x'',\mathbf x',s) \exp(-W[\mathbf J])
\end{equation}
Where we have introduced the Green's function neglecting endpoint Hamiltonian values:
\begin{equation}
K_h(\mathbf x'',\mathbf x',s) =\left(\frac{1}{2\pi}\right)^{D/2}\sqrt{\det\left(\frac{1}{2}\underline \kappa\text{csch}\left(\underline \kappa s\right)\right)}\exp\left(-S_{cl}\right)
\end{equation}

where $S_{cl}$ is the same classical action appearing in the previous section, that of a $J=0$ harmonic system evaluated along the classical path $\mathbf x_{cl}(s')$. The Schwinger functional, which accounts for the terms that depend on $J$ is given by:

\begin{equation}
    W[J]=\int \mathrm d s \mathbf J(s).\mathbf x_{cl}(s) + \frac{1}{2}\int \mathrm d s\int \mathrm d s' \mathbf J(s).\underline G_D(s,s').\mathbf \underline \mathbf J(s')
\end{equation}
Where $G_D$ is (unfortunately) usually also called a many-body Green's function, which for a harmonic system is given by:
\begin{align}
\underline G_D(s,s') = \frac{1}{2}\underline \kappa^{-1}\text{csch}(\underline \kappa s ) \bigg(&\theta \left(s-s'\right) \left(\cosh \left(\underline \kappa \left(s
   -s'-s\right)\right)-\cosh \left(\underline \kappa \left(s +s'-s\right)\right)\right)\\ \nonumber &+\theta
   \left(s'-s\right) \left(\cosh \left(\underline \kappa \left(s -s'-s\right)\right)-\cosh \left(\underline \kappa
   \left(s -s'+s\right)\right)\right)\bigg)
\end{align}

We can calculate an arbitrary perturbation accounting for a potential term going as $V$ according to the formula:
\begin{equation}
    G(\mathbf x'',\mathbf x') =e^{\frac{\beta}{2}\left(H(\mathbf x'')-H(\mathbf x')\right) }\int_0^{\infty}\mathrm d s K_h(\mathbf x'',\mathbf x',s) \exp\left[\int \mathrm d s' V\left(\frac{\delta}{\delta \mathbf J(s')}\right)\right]\exp(-W[\mathbf J])\big|_{\mathbf J=0}
\end{equation}

splitting the Hamiltonian into harmonic and anharmonic terms gives us the expression in the main text, which goes as:
\begin{equation}
    G(\mathbf x'',\mathbf x') =e^{\frac{\beta}{2}\left(V(\mathbf x'')-V(\mathbf x')\right) }\int_0^{\infty}\mathrm d s G_h(\mathbf x'',\mathbf x',s) \exp\left[\int \mathrm d s' V_a\left(\frac{\delta}{\delta \mathbf J(s')}\right)\right]\exp(-W[\mathbf J])\big|_{\mathbf J=0}
\end{equation}

Where we already know the mechanical trick of repetively applying $G_h(\mathbf x'',\mathbf x')$ to any function $\rho$, as was expressed in the previous section. The term $V_a$ can potentially contain terms going as order $x^2$, but it's best to keep them in the perturbative expansion so that $G_h$ can be applied simply each time. 
\subsection{Application to active brownian particle in anharmonic potential}

We can calculate the application of the above for the active anharmonic system described in the main text. Such a system has $\rho(\mathbf x)=\beta^2 a q^{(3)} x^3 A$ where the full state $\mathbf x=(x,A)$ For brevity let us define $q=q^{(3)}$. For ease of nomenclature, we shall define the integral over the harmonic green's function as an operator $\hat O$:
\begin{equation}
    \int_0^{\infty} \mathrm d s\int \mathrm d \mathbf x' G_h(\mathbf x,\mathbf x',s)\rho(\mathbf x') = \hat O \rho
\end{equation}
for this system $V(\mathbf x)=\frac{1}{4}q x^4$ and $V_a(\mathbf x) = 3 q x^2+ q^2 x^6$

The first terms in the anharmonic series then go as:
\begin{equation} \hat {O}[ \rho] + \hat O \left[\beta/2\right(V(\mathbf x)-V(\mathbf x'))\rho] + \hat O\left[\left(\int \mathrm d s'  V_a\left(\frac{\delta}{\delta \mathbf J(s')}\right)\exp(-W[\mathbf J])\big|_{\mathbf J=0}\right)\rho\right] \end{equation}
 the action of an operator $B_{ij}\partial_i\partial_j$ on the term $\exp(-W[J])$ is $\int_0^{s} \mathrm d s'\text{Tr}(\underline G_D(s',s').\underline B)+\frac{1}{2}\mathbf x_{cl}(s').\underline B.\mathbf x_{cl}(s')$. Using the mechanical rules, each of these terms can be evaluated, for now let us see how terms going as $\mathbf x^2$ look:
 \begin{align}
    \hat O \rho &= \frac{3 a q \tau_a^2}{\gamma ^2}  x A \\
    \hat O (\beta/2)V(\mathbf x)\rho) &\sim \mathcal{O}(x^4) \text{ as $\hat{O}$ doesn't act on $x$ and $V\sim x^4$} \\
    \hat O (\beta/2)V(\mathbf x')\rho) &=  \frac{630 a q^2 \tau_a^4}{\beta \gamma ^4}  x A
\end{align}
and to evaluate the Green's terms we use the fact that $\int_0^{s} \mathrm d s'\text{Tr}(\underline G_D(s',s').\underline B) = -(1/2) q s^2 \beta $ where in this example $B=\begin{bmatrix}&3 q,0 \\ &0,0 \end{bmatrix}$. The classical path, which depends on both $\mathbf x'$ and $\mathbf x''$ is given in \eqref{eq:classicalpath}. Taken together, these can be evaluated, yielding:
\begin{align}
    \hat O \left(\int_0^{s} \mathrm d s'\text{Tr}(\underline G_D(s',s').\underline B)\rho \right)&= \frac{18 a q^2 \tau_a^4}{ \beta \gamma ^4}  x A\\
    \hat O \left(\int_0^{s}  \mathrm d s'\frac{1}{2}\mathbf x_{cl}(s').\underline B.\mathbf x_{cl}(s') \rho \right)&= \frac{432 a q^2 \tau_a^4}{\beta \gamma ^4}  x A\\
\end{align}
From which the subequent structure of the series seen in the main text becomes apparent. A similar process can be repeated for terms going as $\sim \mathbf x^4$ etc.

However, a problem arises when the anharmonic series diverges (as is usually the case). In such a scenario, it is better not to perform a series expansion and instead calculate the following integrals directly, which leads to the anharmonic integral and anharmonic integral with dissipative correction approximations respectively:
\begin{equation} \label{eq:anharmintegral}
\int_0^{\infty} \mathrm d s\int \mathrm d \mathbf x' e^{(\beta/2)(V(\mathbf x)-V(\mathbf x'))}G_h(\mathbf x,\mathbf x',s)\rho(\mathbf x')
\end{equation}
\begin{equation}\label{eq:anharmintegralwithdisco}
\int_0^{\infty} \mathrm d s\int \mathrm d \mathbf x' e^{(\beta/2)(V(\mathbf x)-V(\mathbf x'))}G_h(\mathbf x,\mathbf x',s)\rho(\mathbf x')\left(1+\int_0^{s} \mathrm d s' \left[\text{Tr}(\underline G_D(s',s').\underline B))+\frac{1}{2}\mathbf x_{cl}(s',\mathbf x,\mathbf x').\underline B.\mathbf x_{cl}(s',\mathbf x,\mathbf x')\right]\right)
\end{equation}
both of these integrals can be performed "analytically" for an active brownian system with the help of symbolic software such as Mathematica, however the resulting equations are too long to be reproduced here, but are available upon request.

\section{Mode space} \label{app:appD}

Some generic features of the modification of equilibrium systems to non-equilibrium perturbations can be discerned if we analyze the microscopic probability in terms of the modes. This is easily done for a harmonic system. Recalling that the log probability in these systems goes as:

    \begin{equation} \label{eq:logP}
        \log(P(\mathbf x)) = - \frac{1}{2} \beta \mathbf x.\underline \kappa.\mathbf x -\frac{1}{2} \beta^2 \mathbf x .         \int_0^\infty \mathrm d s  \exp(-\beta\underline \kappa s). \underline C .\exp(-\beta\underline \kappa s). \mathbf x + c_1
    \end{equation} 
Where $\underline C$ is some dissipation matrix and $c_1$ is a normalization constant. 

The matrix $\underline \kappa$ will have certain properties. For the system to be conservative in the absence of any non-equilibrium perturbation, the matrix $\kappa$ must be symmetric $\underline \kappa=\underline \kappa^{T}$. A symmetric matrix can always be diagonalized $\underline \kappa=\underline Q^{-1} \underline L \underline Q$ where the matrix $\underline L$ will be a diagonal matrix with each diagonal entry being equal to one of the eigenvalues of the original matrix $L_{ii}=\lambda_i$ and the matrix $\underline Q$ is the matrix of eigenvectors. The matrix exponential for diagonalizable matrices is also well known to have the simplification $e^{\underline \kappa} = \underline Q^{-1}e^{\underline L}\underline Q$. The matrix $Q$ defines a linear transformation, therefore we can define a new space for our problem given by $  \mathbf{ \tilde x} = \underline Q \mathbf x$. Each entry in the vector $\tilde {\mathbf x}$ is a mode of the equilibrium system. This allows re-expression of eq. \ref{eq:logP} as:
\begin{equation}
    \log(P(\mathbf{\tilde x})) = - \frac{1}{2} \beta \mathbf{\tilde x}. \underline L.\mathbf{\tilde x} -\frac{1}{2} \beta^2 \mathbf{\tilde x} .         \int_0^\infty \mathrm d t  \exp(-\beta\underline L t). \underline{\tilde C} .\exp(-\beta\underline L t). \mathbf{\tilde x}
\end{equation}
where $\underline{\tilde C}=\underline Q.\underline C.\underline Q^{-1}$. In mode space, the integral over all path lengths can be performed, leaving the final probability as:
\begin{equation} \label{eq:modespace}
    \log(P(\mathbf{\tilde x})) = - \frac{1}{2} \beta \mathbf{\tilde x}. \underline L.\mathbf{\tilde x} -\frac{1}{2}  \beta^2 \mathbf{\tilde x} .        \underline{\tilde{D}}. \mathbf{\tilde x}
\end{equation}
where we have a new matrix $\underline{\tilde D}$ whose elements are given by:
\begin{equation} \label{eq:D}
    \tilde D_{ij}=\frac{\tilde C_{ij}}{\lambda_i+\lambda_j}
\end{equation}

As the matrix $C$ for driving under a non-conservative force goes as $\underline C=\underline \kappa.\underline M = \underline Q^{-1}.\underline L.\underline Q. \underline M$, which then means $\tilde{\underline C}=\underline L \underline Q.\underline M.\underline Q^{-1}$ and defining $\tilde{\underline M}=\underline Q.\underline M.\underline Q^{-1}$
\begin{equation} \label{eq:D2}
    \tilde D_{ij}=\frac{\lambda_i \tilde M_{ij}}{\lambda_i+\lambda_j}
\end{equation}
we get the property that the non-conservative force affects the variance of each mode by an amount that is roughly independent of eigenvalue of the mode.

For an active polymer, half of each "modes" refers to the degrees of freedom of the active forces $A$. When integrating out over these degrees of freedom, we can obtain the effective $1/(1+\lambda^2)$ expression that is seen in the main text. The true expressions can be found analytically, but are rather long.

\section{Simuation details}

 The simulated harmonic system of $N$ particles where the form of potential between particles $i$ and $j$ is given by:
\begin{equation} \label{eq:pot}
\phi(\mathbf x_i,\mathbf x_j) = \frac{1}{2}k(\mathbf x_i-\mathbf x_j)^2
\end{equation}
where $k$ is the spring constant. For simplicity, we consider a system where particles are only bound to their 1D neighbors, leading to a polymer. The system is thermalized through the Langevin thermostat, where the noise is uncorrelated in time:
\begin{equation}
    \langle \boldsymbol \xi_i(t) \boldsymbol \xi_i(t') \rangle = 2 k_b T_i \gamma \delta(t-t')
\end{equation}
where $k_b T$ is the temperature. For the cases where an anharmonic potential is introduced, we modify the relationship in ~\eqref{eq:pot} to 
\begin{equation} \label{eq:pot}
\Phi(\mathbf x_i,\mathbf x_j) = \frac{1}{2}k(\mathbf x_i-\mathbf x_j)^2+\frac{1}{4}q(\mathbf x_i-\mathbf x_j)^4
\end{equation}

and we integrate numerically the following equations
\begin{align}
    \gamma \frac{\mathrm d \mathbf x_i}{\mathrm d t} &= \mathbf F_i(\{\mathbf x\})+ a \mathbf A_i + \boldsymbol \xi_i(t) \\    
    \gamma \frac{\mathrm d \mathbf A_i}{\mathrm d t} &= -(\gamma/\tau_a)\mathbf A_i + \boldsymbol \xi_i(t) \\
\end{align}
    where $\mathbf F_i(\{\mathbf x\})=-\nabla_i \sum_j^{N-1} \Phi(\mathbf x_j,\mathbf x_{j+1}) $

\end{document}